\documentclass[conference]{IEEEtran}
\IEEEoverridecommandlockouts
\usepackage{cite}
\usepackage{amsmath,amssymb,amsfonts}
\usepackage{algorithmic}
\usepackage{graphicx}
\usepackage{textcomp}
\usepackage{xcolor}
\usepackage{url}
\usepackage{setspace}
\usepackage{subcaption}
\usepackage{subfig}
\usepackage{svg}
\usepackage{tikz}
\IEEEoverridecommandlockouts
\newcommand\copyrighttext{%
  \footnotesize \textcopyright 2026 IEEE.  Personal use of this material is permitted.  Permission from IEEE must be obtained for all other uses, in any current or future media, including reprinting/republishing this material for advertising or promotional purposes, creating new collective works, for resale or redistribution to servers or lists, or reuse of any copyrighted component of this work in other works.}
\newcommand\copyrightnotice{%
\begin{tikzpicture}[remember picture,overlay]
\node[anchor=south,yshift=10pt] at (current page.south) {\fbox{\parbox{\dimexpr\textwidth-\fboxsep-\fboxrule\relax}{\copyrighttext}}};
\end{tikzpicture}%
}
\begin{document}

\title{Performance Comparison of 5G NR Uplink MIMO and Uplink Carrier Aggregations on Commercial Network}

\author{\IEEEauthorblockN{Henry Shao}
\IEEEauthorblockA{
\textit{Sylvania Southview High School}\\
Sylvania, OH, USA\\
hshao01@proton.me\\
\vspace{-8mm}}
\and
\IEEEauthorblockN{Kasidis Arunruangsirilert}
\IEEEauthorblockA{\textit{Department of Computer Science and Communications Engineering} \\
\textit{Waseda University}\\
Tokyo, Japan \\
kasidis@fuji.waseda.jp\\}
\vspace{-8mm}
}

\maketitle
 \copyrightnotice
\setstretch{0.96}
\begin{abstract}
%Demands for uplink on mobile networks are increasing with the rapid development of social media platforms, 4K/8K content creation, IoT applications, and Fixed Wireless Access (FWA) broadband. As a result, Uplink MIMO (UL-MIMO) and Uplink Carrier Aggregation (UL-CA) have been widely deployed for the first time on commercial 5G networks. UL-MIMO enables the transmission of two data streams in strong RF conditions, theoretically doubling throughput and efficiency. In challenging conditions, one layer is transmitted over two antennae to increase reliability. On the other hand, UL-CA allows for simultaneous upload on greater channel widths, allowing more resources to be assigned to a single UE for higher throughput.

Demands for uplink on mobile networks are increasing with the rapid development of social media platforms, 4K/8K content creation, IoT applications, and Fixed Wireless Access (FWA) broadband. As a result, Uplink MIMO (UL-MIMO) and Uplink Carrier Aggregation (UL-CA) have been widely deployed for the first time on commercial 5G networks. UL-MIMO enables the transmission of two data streams on one frequency band in strong RF conditions, theoretically doubling throughput and efficiency. On the other hand, UL-CA allows for simultaneous upload on greater channel widths, allowing more resources to be assigned to a single UE for higher throughput.

%In the United States, T-Mobile USA, a mobile network operator (MNO), has deployed network-wide 5G Standalone (SA), along with UL-MIMO on Time Division Duplex (TDD) band n41 and UL-CA between TDD and Frequency Division Duplex (FDD) NR bands. In this paper, the uplink throughput performance of UL-MIMO and UL-CA will be evaluated on the commercial T-Mobile 5G network on routes in rural, suburban, and urban areas, capturing a variety of RF environments and modes of transportation. It was found that, even with the efficiency gains, UL-MIMO  yields slower uplink throughput in most scenarios. However, in stronger RF conditions, such as in urban areas, UL-MIMO can provide an adequate user experience, so capacity can be conserved by reserving UL-CA for UE in worse RF conditions. 

In the United States, T-Mobile USA, a mobile network operator (MNO), has deployed network-wide 5G Standalone (SA), along with UL-MIMO on Time Division Duplex (TDD) band n41 and UL-CA between TDD and Frequency Division Duplex (FDD) NR bands. In this paper, the uplink throughput performance of UL-MIMO and UL-CA will be evaluated on the commercial T-Mobile 5G network on a variety of RF environments and modes of transportation. It was found that, even with the efficiency gains, UL-MIMO yields slower uplink throughput in most scenarios. However, in stronger RF conditions, UL-MIMO can provide an adequate user experience, so capacity can be conserved by reserving UL-CA for UE in weaker RF conditions. 

\end{abstract}

\setstretch{0.95}
\section{Introduction}
  %  In the 4G era, the primary use cases for Radio Access Network (RAN) were web browsing and simple communication (messaging, E-Mails, etc.), which are primarily downlink-heavy applications. Mobile Network Operators (MNOs) around the world were focused on improving download performance and capacity, which is limited in 3G Universal Mobile Telecommunications System (UMTS) deployments. With the introduction of Downlink MIMO (DL-MIMO) and Time-Division Duplexing LTE (TDD-LTE), networks continued to expand downlink capacity as traffic-heavy applications such as multimedia content consumption and cloud storage became prevalent. Carrier Aggregation (CA) was defined with LTE Advanced (LTE-A) in 3GPP TS 36.300 Release 11\cite{3GPP_36-300}, where multiple component carriers (CCs) could be aggregated in both downlink and uplink to improve single-user throughput. While CA is commonly deployed worldwide for downlink, it is seldom supported in the uplink direction, both by MNOs and User Equipment (UE) manufacturers. At the time, most uplink-heavy tasks were completed over fixed broadband connections, rather than over mobile networks. 

    In the 4G era, mobile Network Operators (MNOs) around the world were focused on improving download performance and capacity, which was limited in 3G Universal Mobile Telecommunications System (UMTS) deployments. Carrier Aggregation (CA) was defined with LTE Advanced (LTE-A) in 3GPP TS 36.300 Release 11\cite{3GPP_36-300}, where multiple component carriers (CCs) could be aggregated in both downlink and uplink to improve single-user throughput. While CA is commonly deployed worldwide for downlink, it is seldom supported in the uplink direction on 4G, both by MNOs and User Equipment (UE) manufacturers. At the time, most uplink-heavy tasks were completed over fixed broadband connections, rather than over mobile networks.

In the late 2010s, the widespread adoption of smartphone systems-on-chips (SoCs) with efficient 4K video encoding, such as High Efficiency Video Coding (HEVC)\cite{qualcomm_855}, enabled the creation of high-quality mobile content. Around the same time, TikTok \cite{tiktok_2025}, Facebook \cite{constine_2016}, and Instagram \cite{constine_2015} introduced live streaming on their platforms, encouraging on-the-go content creation. In addition to the rise of real-time video conferencing due to COVID-19, emerging telepresence technologies also require a high-throughput and reliable uplink connection for an optimal user experience \cite{orduna_gonzalez-sosa_boudouraki_elmimouni_perez_gutierrez_ahumada-newhart_cesar_2025}. 
    
    %With the introduction of 5G New Radio (NR), new uplink heavy use cases were also introduced, including Ultra High Definition (UHD) 4K/8K livestreaming, Self-Driving Vehicles, Virtual Reality (VR)/Augmented Reality (AR) content live streaming, Internet of Things (IoT), and Industrial Automation\cite{erunkulu20215g}. Furthermore, with the rapidly increasing adoption of Fixed Wireless Access (FWA) broadband over mobile 5G networks\cite{morris_2025}, customers expect similar uplink throughput and reliability to fixed broadband to support large file transfers and real-time applications.  With the rise in uplink traffic carried over 5G networks\cite{ericsson_2025}, efficient use of limited midband spectrum is crucial\cite{gsma_2025}. To meet these demands, Uplink Multiple-Input Multiple-Output (UL-MIMO) and Uplink Carrier Aggregation (UL-CA) have been widely adopted for the first time by MNOs on 5G networks, as well as UE manufacturers. UL-MIMO increases uplink efficiency by taking advantage of two transmit (Tx) streams compared to just one \cite{10118777}, while UL-CA transmits over two CCs with one Tx stream per carrier, totaling two Tx streams. This creates a healthy ecosystem of Mobile Network and UE that supports advanced 5G uplink features, which enhance Quality of Experience (QoE) in uplink-heavy use cases.
\begin{figure}[!tpb]
    \centering
    \includesvg[width=0.37\textwidth,inkscapelatex=false]{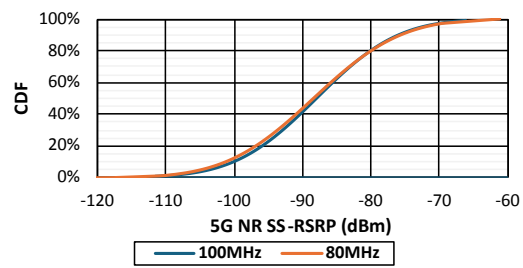}
    \caption{CDF of SS-RSRP of 100 MHz and 80 MHz Channels of Band n41}
    \vspace{-7.5mm}
    \label{fig:n41_cdf}
    
\end{figure}
    Uplink-heavy use cases were introduced along with 5G New Radio (NR), including Self-Driving Vehicles, Virtual Reality (VR)/Augmented Reality (AR) content live streaming, and Industrial Automation\cite{erunkulu20215g}. Furthermore, with the rapidly increasing adoption of Fixed Wireless Access (FWA) broadband over mobile 5G networks\cite{morris_2025}, customers expect uplink throughput and reliability similar to fixed broadband.  With the rise in uplink traffic carried over 5G networks\cite{ericsson_2025}, efficient use of limited mid-band spectrum is crucial\cite{gsma_2025}. To meet these demands, Uplink Multiple-Input Multiple-Output (UL-MIMO) and Uplink Carrier Aggregation (UL-CA) have been widely adopted for the first time by MNOs on 5G networks, as well as UE manufacturers. UL-MIMO increases uplink efficiency by taking advantage of two transmit (Tx) streams compared to just one \cite{10118777}, while UL-CA transmits over two CCs with one Tx stream per carrier, totaling two Tx streams. This creates a healthy ecosystem of Mobile Network and UE that supports advanced 5G uplink features, which enhance Quality of Experience (QoE) in uplink-heavy use cases.

 %Combining UL-MIMO and UL-CA is also possible as defined in 3GPP TS 38.214 Release 16, which can be referred to as UL Tx switching\cite{3GPP_38-214}; however, this is rarely supported on handsets and networks at the time of study. This forces operators to choose between UL-MIMO or UL-CA, each of which performs better in specific RF conditions. Therefore, it is critical for operators to understand which uplink enhancement strategy yields better results under what real-world conditions, as they are currently mutually exclusive options. In this paper, the real-world performance of UL-MIMO will be compared to the performance of UL-CA on a live 5G NR Standalone (SA) network. Drive and walk tests performed on rural, suburban, and dense urban routes in the metropolitan area of Toledo, Ohio, are the main point of discussion, as they encompass a variety of network conditions and common methods of transportation in the United States. This paper is organized as follows: Section \ref{sec_experiment} will discuss the experiment environment; Section \ref{sec_result} will provide results and an analysis of the results; finally, Section \ref{sec_conclusion} will contain the conclusion and a discussion of future works.

    Combining UL-MIMO and UL-CA is also possible as defined in 3GPP TS 38.214 Release 16, which can be referred to as UL Tx switching\cite{3GPP_38-214}; however, this is rarely supported on handsets and networks at the time of this study. Therefore, it is critical for operators to understand which uplink enhancement strategy yields better results under what real-world conditions. In this paper, the real-world performance of UL-MIMO will be compared to the performance of UL-CA on a live 5G NR Standalone (SA) network. Drive and walk tests performed on rural, suburban, and dense urban routes in the metropolitan area of Toledo, Ohio are the main point of discussion, as they encompass a variety of network conditions and common methods of transportation in the United States. This paper is organized as follows: Section \ref{sec_experiment} will discuss the experiment environment; Section \ref{sec_result} will provide results and an analysis of the results; finally, Section \ref{sec_conclusion} will contain the conclusion and a discussion of future works.

\section{Experiment Environment} \label{sec_experiment}

\subsection{Network Environment}

This analysis required a 5G Standalone (SA) network supporting both UL-MIMO and UL-CA. In the United States, the T-Mobile USA network meets these criteria. The network utilizes two carriers in band n41: the first carrier with a 100 MHz bandwidth (2.51–2.61 GHz) and the second carrier with an 80 MHz bandwidth (2.61–2.69 GHz). Figure \ref{fig:n41_cdf} shows that the difference in signal propagation between these channels was negligible. This is likely because both carriers are transmitted from the same Active Antenna Unit (AAU) installed on the same base stations, resulting in similar measured Reference Signal Received Power (RSRP) and Signal-to-Interference-and-Noise Ratio (SINR).

%Because locking the UE to a specific carrier would disable UL-CA, and since both n41 carriers occupy the same 5G frequency band, it was not possible to force the UE onto the 100 MHz carrier using an NR-ARFCN or frequency band lock. Therefore, data was collected from a mixture of 100 MHz and 80 MHz channels during field tests. To mitigate performance discrepancies arising from different channel bandwidths, all throughput measurements were normalized prior to analysis. This normalization was based on the proportion of scheduled Resource Blocks (RBs) relative to the maximum RB count of a 100 MHz carrier. This method also serves as a way to normalize the variations in cell load for the other frequency bands. T-Mobile's network also deploys two 20 MHz Frequency-Division Duplex (FDD) bands: n25 (1985–2005 MHz uplink) and n71 (622–642 MHz uplink).

Because locking the UE to a specific carrier would disable UL-CA, and since both n41 carriers occupy the same 5G frequency band, it was not possible to force the UE onto the 100 MHz carrier using an NR-ARFCN or frequency band lock. Therefore, data was collected from a mixture of 100 MHz and 80 MHz channels during field tests. To mitigate performance discrepancies arising from different channel bandwidths, all throughput measurements were normalized prior to analysis, which is discussed below. T-Mobile's network also deploys two 20 MHz Frequency-Division Duplex (FDD) bands: n25 (1985–2005 MHz uplink) and n71 (622–642 MHz uplink).

%For the network configuration, UL-MIMO is enabled network-wide on the n41 band on both carriers, and UL-CA enabled between FDD and TDD bands. Uplink experiments was conducted on n41 for gathering data on UL-MIMO performance, while data will be gathered for each n41 + n25 and n41 + n71 to gather data on UL-CA performance. The UE was able to freely switch between n41 or the respective FDD band as Primary Component Carrier (PCC). Over 90\% of data points during UL-CA trials recorded n41 as the PCC, so most of the evaluation will be performed on data points containing n41 as the PCC; however, some of the data points where n41 was recorded as a Secondary Component Carrier (SCC) will also be included for reference purposes.

The network was configured with UL-MIMO enabled on both n41 carriers and UL-CA activated between the TDD (n41) and FDD (n25, n71) bands. UL-MIMO performance data was collected from experiments on the n41 band, while UL-CA performance was assessed using data with CA\_n41AA-n25AA and CA\_n41AA-n71AA configurations. Due to the network configuration, the UE was able to dynamically assign either the n41 or the FDD band as the Primary Component Carrier (PCC). However, since n41 served as the PCC in over 90\% of the UL-CA measurements, the evaluation focuses on data points with this configuration for consistency. %Data points where n41 was recorded as a Secondary Component Carrier (SCC) are also included for reference.

%For evaluation accuracy, data points that do not represent the performance of the upload enhancement of interest will be filtered. For example, if the UE drops the SCC during a UL-CA trial, the data points collected when the SCC is absent will be discarded. Data points immediately preceding, succeeding, or during a handoff will also be filtered, as abnormal values, including missing values and irregular throughput values, may skew results.

%All tests were conducted with a T-Mobile "Business Unlimited Ultimate" mobile plan, allowing for unlimited QoS Class Index (QCI) 6 data access. Due to load from other subscribers on the sector being tested, all throughput values for each band were collected were adjusted with the following equation: 

To ensure the accuracy of the evaluation, the data points that were not representative of the specific upload feature under investigation were filtered. During the UL-CA trials, any data points collected when the Secondary Component Carrier (SCC) was inactive were discarded. Measurements recorded immediately preceding, during, or succeeding a network handoff were removed to avoid irregular throughput values. Furthermore, data points where the UE was connected to cell sites lacking all three NR bands were removed to ensure that the analysis would not be affected by the difference in coverage or deployment patterns of each band. \looseness=-1

All tests were conducted using a T-Mobile "Business Unlimited Ultimate" mobile plan, which provides unlimited prioritized data access at QoS Class Index (QCI) 6. As mentioned earlier, to account for network load from other subscribers and differences in channel bandwidth on the tested cell, all collected throughput values were normalized using the following equation:

\vspace{3mm}

%\(AdjustedThroughput = MeasuredThroughput * MaximumRBs / MeasuredRBs\)
\centerline{\(Thpt_{Normalized} = Thpt_{Measured} \cdot \frac{RB_{Max,100M}}{RB_{Measured}}\)}

\vspace{3mm}

Where \(Thpt_{Measured}\) is the measured throughput in Megabits-per-second (Mbps) and \(RB_{Measured}\) is the scheduled Physical Uplink Shared Channel (PUSCH) Resource Block (RB) in that one-second interval of the data point. Whereas \(RB_{Max,100M}\) is the maximum number of RB that can be scheduled to the UE on a 100 MHz channel, accounting for the TDD configuration used by T-Mobile USA (Slot (DL/UL): 7/2, Symbol (DL/UL): 4/4). This yields normalized throughput, \(Thpt_{Normalized}\), in Mbps, which represents the theoretical throughput that would be achieved if all resource blocks (RBs) were assigned to the UE at that given instant.

The result is a normalized theoretical throughput that would be achieved if all resource blocks (RBs) were assigned to the UE at that given time, with the difference in channel bandwidth and cell load normalized. From the calculation, the maximum RBs available on a 100 MHz channel of n41 with 30 kHz Subcarrier Spacing (SCS) is \(457 * 273 = 124761\), and the maximum RBs for a 20 MHz channel of both n71 and n25 with SCS of 15 kHz is \(1000 * 106 = 106000\) each. The normalized throughput for each band during the UL-CA trials was then summed up to obtain an adjusted total throughput measurement.

\subsection{User Equipment (UE) and Data Collection}

% For UE, a Samsung Galaxy S24 (SM-S921U1) equipped with the Snapdragon X75 Modem was used during the rural and suburban drive tests. This device supports UL-CA between TDD band n41 and FDD bands n25 and n71. UL-MIMO was not enabled by default, but is already supported by the hardware on band n41, so the modem firmware was modified to enable UL-MIMO for the corresponding trials. However, for UL-CA trials, UL-MIMO was disabled by the same method, and only the corresponding bands for each trial were enabled through the built-in Band Selection menu. Data were recorded using \textit{AirScreen}, a software utility that collects in-depth information about the network connection, including RF parameters, throughput, and signaling. Data along suburban and rural test routes was collected by connecting the Samsung Galaxy S24 with modem diagnostic ports exposed with Android Debug Bridge (ADB) enabled via USB to a Windows 11 laptop equipped with \textit{AirScreen}.  

For suburban and rural drive tests, a Samsung Galaxy S24 (SM-S921U1) equipped with a Snapdragon X75 Modem was utilized as the UE. This device supports UL-CA between TDD band n41 and FDD bands n25 and n71. For trials involving UL-MIMO, which is not enabled by default on band n41, the modem firmware was modified to activate the feature. Conversely, for UL-CA trials, UL-MIMO was disabled through a similar modification, and specific bands were isolated using the UE's native Band Selection menu. The UE was connected to a laptop to record RF parameters, throughput, and signaling data using the \textit{AirScreen} software.

An alternative UE, Pixel 9 Pro (GR83Y) with an Exynos 5400 modem, was used for the urban walk test. This device supports UL-CA and has UL-MIMO enabled by default. To enable data collection, root access was obtained for the use of \textit{Network Signal Guru (NSG)}, a similar logging utility that runs directly on the smartphone. The use of this device also provided an opportunity to assess the performance of UL-MIMO and UL-CA on both Qualcomm and Samsung modems. To ensure consistent testing conditions, band selection was enforced by replacing the stock band combination file (.binarypb) with a custom file containing only the combinations of interest for each trial. %This method permitted the use of a single device for all urban tests, eliminating potential variables from differences in antenna design or RF circuitry.

Uplink throughput was measured during continuous upload stress tests using the scripting feature in both \textit{AirScreen} and \textit{NSG}. The test script executed a continuous stress test via HTTP POST to a test server. The network was observed to be configured to release the SCC during periods of low downlink traffic. To prevent this, a continuous iperf3 downlink stream, rate-limited to 1 Mbps, was run concurrently with all uplink tests. This procedure was applied uniformly across both UL-MIMO and UL-CA trials to ensure consistency. \looseness=-1
\begin{figure}[!tbp]
    \centering
    \begin{subfigure}[t]{0.156\textwidth}
        \centering
        \includegraphics[width=0.97\linewidth]{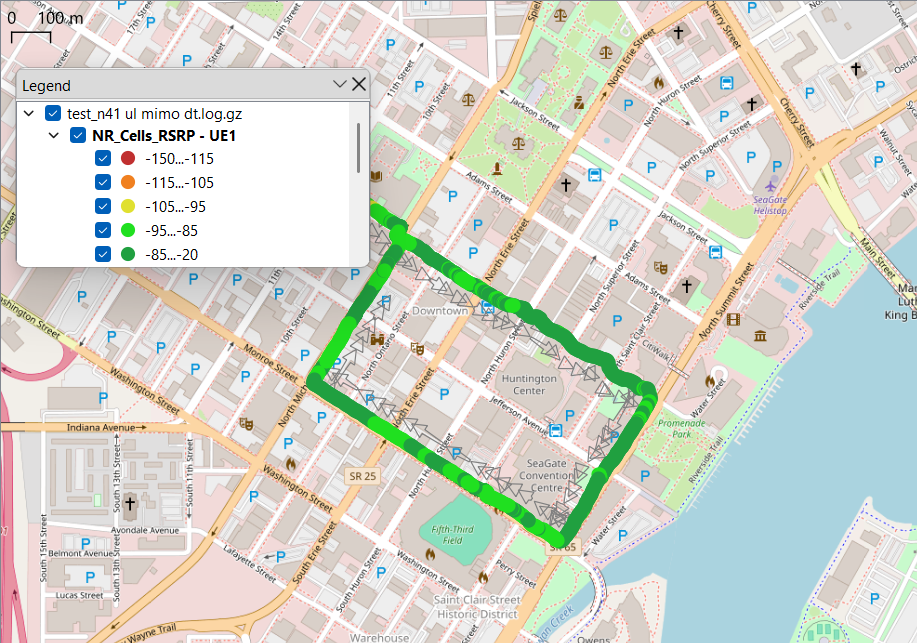}
        \caption{Urban Route}
        \label{fig_urban}
    \end{subfigure}%
    \begin{subfigure}[t]{0.169\textwidth}
        \centering
       \includegraphics[width=0.97\linewidth]{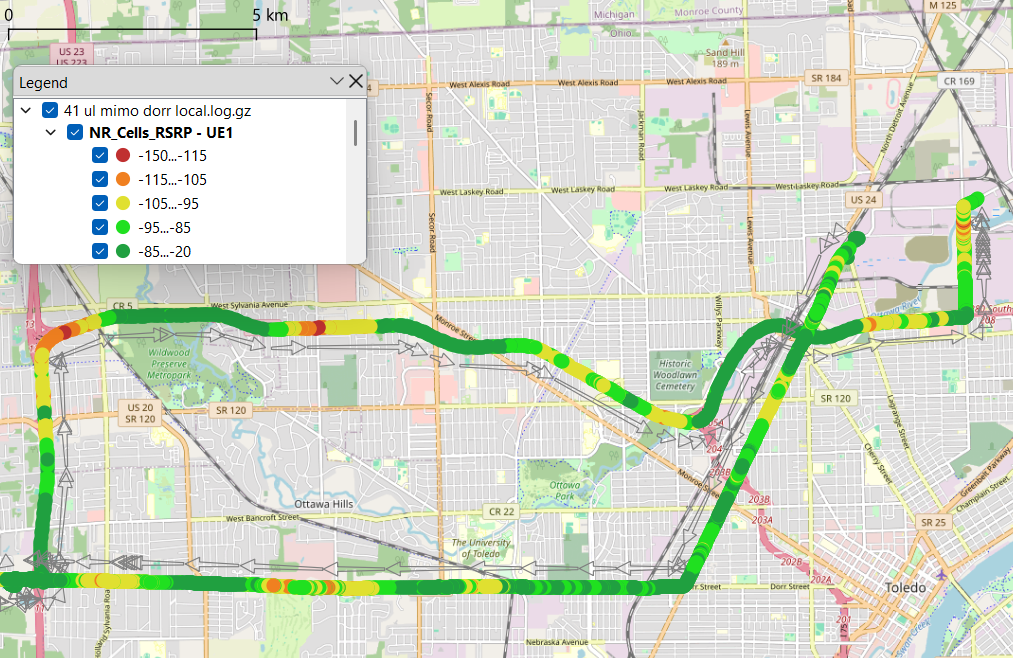}
        \caption{Suburban Route}
        \label{fig_suburban}
    \end{subfigure}%
     \begin{subfigure}[t]{0.157\textwidth}
        \centering
        \includegraphics[width=0.97\linewidth]{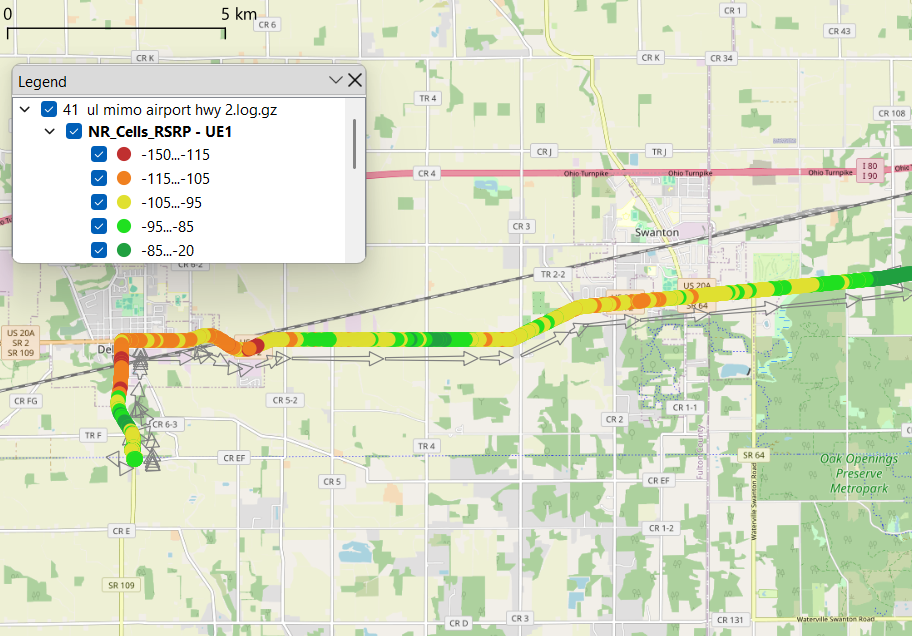}
        \caption{Rural Route}
        \label{fig_rural}
    \end{subfigure}
    \captionsetup{justification=centering}
    \caption{Map of Test Routes. Color shows 5G NR SS-RSRP.}
    \vspace{-5mm}
    \label{route_maps}
\end{figure}

\begin{figure}[!tbp]
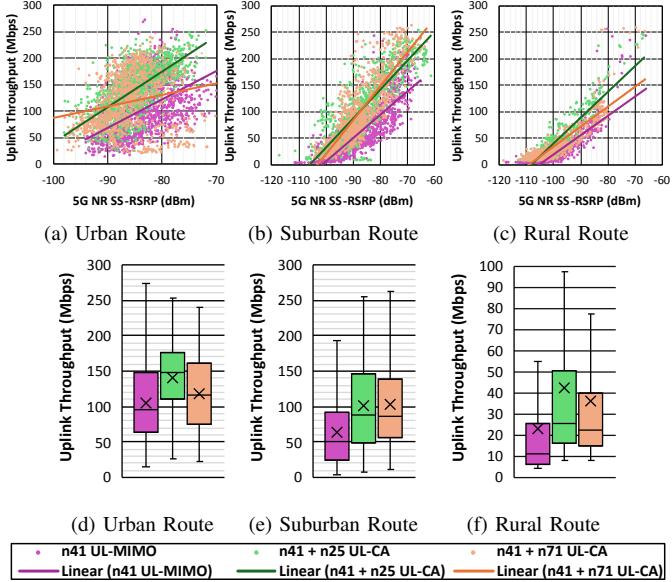

    \centering
    \begin{subfigure}[t]{0.16\textwidth}
        \centering
        \includesvg[width=\linewidth,inkscapelatex=false]{urban_graph_new}
        \caption{Urban Route}
        \label{urban_graph}
    \end{subfigure}%
    \begin{subfigure}[t]{0.16\textwidth}
        \centering
        \includesvg[width=\linewidth,inkscapelatex=false]{suburban_graph_new}
        \caption{Suburban Route}
        \label{suburban_graph}
    \end{subfigure}
     \begin{subfigure}[t]{0.16\textwidth}
        \centering
        \includesvg[width=\linewidth,inkscapelatex=false]{rural_graph_new}
        \caption{Rural Route}
        \label{rural_graph}
    \end{subfigure}
    \\
    \begin{subfigure}[t]{0.14\textwidth}
        \centering
        \includesvg[width=\linewidth,inkscapelatex=false]{urban_dist_crop}
        \caption{Urban Route}
        \label{urban_dist_crop}
    \end{subfigure}%
    \begin{subfigure}[t]{0.14\textwidth}
        \centering
        \includesvg[width=\linewidth,inkscapelatex=false]{suburban_dist_crop}
        \caption{Suburban Route}
        \label{suburban_dist_crop}
    \end{subfigure}
     \begin{subfigure}[t]{0.14\textwidth}
        \centering
        \includesvg[width=\linewidth,inkscapelatex=false]{rural_dist_crop}
        \caption{Rural Route}
        \label{rural_dist_crop}
    \end{subfigure}\\
    \vspace{0.7mm}
    \begin{subfigure}[t]{0.48\textwidth}
        \centering
        \includesvg[width=\linewidth,inkscapelatex=false]{legend}
    \end{subfigure}
    
    % \captionsetup{justification=left}
    \caption{ \textbf{Field Test Result}:  Uplink throughput performance by Configuration and Test Route. Sub-figures (a), (b), and (c) plot the measured Uplink Throughput (Mbps) vs 5G NR SS-RSRP (dBm) for the urban, suburban, and rural routes, respectively. Sub-figures (d), (e), and (f) show the box plots summarizing the distribution of Uplink Throughput (Mbps) for each configuration across the three test environments.}
    \label{uplink_distributions}
    \vspace{-6mm}
\end{figure}
\subsection{Test Routes}

Three representative test routes were chosen in the Toledo, Ohio region where either walk tests or drive tests were conducted. Uplink throughput measurements were collected for n41 UL-MIMO, CA\_n41AA-n25AA, and CA\_n41AA-n71AA. The test routes are as follows:
\begin{itemize}
    \item \textbf{Urban Route}: This route in the downtown core passed some of the most popular attractions in the area, including convention centers and stadiums (see Figure \ref{fig_urban}), where uplink-heavy applications, such as video calling and social media uploads, are commonly utilized. The mode of transportation for this route is walking, as this urban environment experiences relatively high amounts of foot traffic. Three sites served this area, providing strong RF conditions for the entire route. 
        %\item This route is located in the dense downtown of Toledo, Ohio, and passes some of the most popular attractions in the area. This route begins at the Toledo Lucas County Main Library and passes stadiums, parks, and convention centers. These are all locations where people commonly use their phones for uplink heavy applications like social media and video calling. The mode of transportation for this route is walking, as this urban environment experiences relatively high amounts of foot traffic. Three sites serve this area, providing strong RF conditions for the entire route. 
    \item \textbf{Suburban Route}: This route included travel on the freeway from exit 15 on Interstate 475 eastward to exit 207 on Interstate 75 (see Figure \ref{fig_suburban}). Travel on local roads included N Detroit Ave. and Dorr St., passing through residential neighborhoods, recreational spaces, and commercial areas. This route represents a typical 5G network deployment in suburban environments.  
    \item \textbf{Rural Route}: Traveling between Swanton, OH and Delta, OH on Ohio State Route 2 and State Route 109, this route passes small towns and farmland (see Figure \ref{fig_rural}). Agricultural IoT and 5G FWA broadband are common use cases that require adequate uplink throughput. The serving sites are further away for most of the route, therefore, capturing uplink performance in weaker RF conditions. \looseness=-1

\end{itemize}

\section{Results and Analysis} \label{sec_result}
\subsection{Physical Uplink Throughput}

\begin{figure}[!tbp]
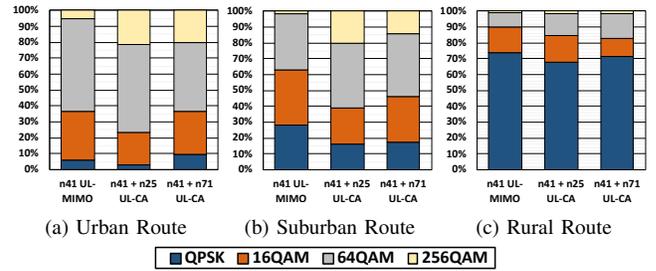

    \centering
    \begin{subfigure}[t]{0.15\textwidth}
        \centering
        \includesvg[width=\linewidth,inkscapelatex=false]{urban_mod}
        \caption{Urban Route}
        \label{urban_mod}
    \end{subfigure}
    \begin{subfigure}[t]{0.15\textwidth}
        \centering
        \includesvg[width=\linewidth,inkscapelatex=false]{suburban_mod}
        \caption{Suburban Route}
        \label{suburban_mod}
    \end{subfigure}
     \begin{subfigure}[t]{0.15\textwidth}
        \centering
        \includesvg[width=\linewidth,inkscapelatex=false]{rural_mod}
        \caption{Rural Route}
        \label{rural_mod}
    \end{subfigure}\\
    \vspace{0.7mm}
    \begin{subfigure}[t]{0.25\textwidth}
        \centering
        \includesvg[width=\linewidth,inkscapelatex=false]{legend_mod}
    \end{subfigure}
    \vspace{-0.5mm}
    % \captionsetup{justification=centering}
    \caption{\textbf{Field Test Result}: Uplink Modulation by Configuration and Test Route}
    \vspace{-8mm}
    \label{uplink_modulations}
\end{figure}

%Figures \ref{urban_graph}, \ref{suburban_graph}, and \ref{rural_graph} shows the relationship between the SS-RSRP (dBm) and the uplink throughput (Mbps) at the Physical Layer (PHY). The results demonstrate that UL-CA configurations consistently achieve higher uplink throughput compared to UL-MIMO across all tested environments. This performance advantage stems from two primary factors. Firstly, UL-CA combines the bandwidth of multiple component carriers (CCs), allowing for more resource blocks to be assigned to a single UE at a time. This demonstrates that wider bandwidths consistently yields a higher performance uplink than smaller bandwidths counterpart (e.g. single channel UL-MIMO only). Secondly, due to the lower uplink carrier frequency of n25 (1.9 GHz) versus n41 (2.5 GHz), less path loss is experienced, which results in a higher uplink performance at the same transmission power \cite{11133825}. 

Figures \ref{urban_graph}, \ref{suburban_graph}, and \ref{rural_graph} plot the Physical Layer (PHY) uplink throughput as a function of SS-RSRP. The results demonstrate that UL-CA configurations consistently outperform UL-MIMO across all tested environments. %This performance advantage is attributable to two primary factors.

The advantages of UL-CA are most pronounced in weak RF conditions. The secondary FDD carriers often benefit from more favorable propagation characteristics \cite{11133825}, enabling the use of a higher MCS index compared to the mid-band TDD carrier. In the urban environment, for example, the 10th percentile throughput of the highest performing UL-CA configuration, CA\_n41AA-n25AA, exceeded that of UL-MIMO by 65.8\% (Figure \ref{urban_dist_crop}). This performance gap widened substantially to 135\% on the rural route (Figure \ref{rural_dist_crop}).

In the suburban tests, cell load characteristics introduced another performance dimension. The 10th percentile performance of CA\_n41AA-n71AA was 39.7\% greater than CA\_n41AA-n25AA and 265\% greater than UL-MIMO (Figure \ref{suburban_dist_crop}). This is attributed to the network's band-steering policy, which heavily prioritizes higher-frequency bands (n41 and n25). Consequently, the 600 MHz band (n71) was less congested. With fewer users on the n71 carrier, there is lower interference and a higher SINR, leading to greater spectral efficiency and throughput for UEs utilizing that band.

%Furthermore,  Due to the weak RF conditions of the rural test route, the UE likely opted to transmit in diversity transmission mode instead of spatial multiplexing mode, resulting in little performance benefit in operating in rank 2 compared to rank 1. UL-CA is favorable in low-density environments for real-time uplink applications, like video calling, and uplink-heavy applications, such as agricultural IoT sensing and fixed wireless broadband. 

From a network resource perspective, UL-MIMO remains a viable strategy, particularly in dense urban environments. The urban walk test showed that n41 UL-MIMO achieved a 10th percentile throughput of 43.4 Mbps and a median of 96.1 Mbps. While UL-CA offers higher single-user throughput, it consumes resources on two carriers. The resources on the supplemental FDD carrier could otherwise be allocated to serve cell-edge users. Therefore, in traffic-heavy scenarios, network operators might steer UEs toward UL-MIMO to maximize overall sector capacity.

Furthermore, the observed performance differences are also influenced by dynamic RF conditions resulting from network load on specific bands. While this analysis normalizes for the number of scheduled Resource Blocks (RBs), it does not account for interference and spectral efficiency degradation caused by other UEs, including those in an RRC Idle state. Therefore, in strong RF conditions where propagation is not the primary constraint, operators can dynamically steer UEs between UL-MIMO and specific UL-CA combinations to maximize spectral efficiency for each scheduled resource. %This strategy, based on real-time, band-specific network usage, would ensure that the spectral efficiency for each scheduled resource is maximized.

\subsection{Modulation and MIMO Performance} \label{sec_MIMO}

The performance of UL-MIMO is fundamentally dependent on the UE's dynamic selection between two transmission modes: a high-throughput spatial multiplexing mode (rank 2) that sends two data streams simultaneously, and a highly reliable transmit diversity mode (rank 1) that sends a single data stream over both antennas for redundancy. This selection is governed by RF conditions; as the signal weakens, the UE reverts to the more robust rank-1 transmission, with a typical threshold around an SS-RSRP of -80 dBm \cite{10118777}. Our trials confirmed this behavior, showing that the single-stream transmit diversity mode was used for 17.5\% of transmissions in the urban test, 59.4\% in the suburban route, and 89.5\% in the rural route. This frequent fallback to a single data stream in suboptimal RF conditions limits the throughput potential of UL-MIMO. \looseness=-1

%A second major constraint arises from power allocation. When operating in rank-2 mode, the UE's total transmission power is split between the two data streams. This reduces the power per stream, making it difficult to achieve the high Signal-to-Noise-and-Interference Ratio (SINR) required for higher-order modulation schemes like 256QAM. This limitation is evident in the modulation usage data (Figure \ref{uplink_modulations}). In the urban test, 256QAM usage for UL-MIMO was only 5.2\%, compared to a range of 18.6-21.6\% for UL-CA configurations (Figure \ref{urban_mod}). The disparity widened in the suburban route, where 256QAM usage during UL-MIMO dropped to 1.4\%, while the CA\_n41AA-n25AA configuration maintained a usage of 20.4\% (Figure \ref{suburban_mod}).  In the rural scenario, however, the RF environment became the overriding constraint, forcing all configurations to rely on the lowest-order modulation, with QPSK being used for over 68\% of Resource Blocks (Figure \ref{rural_mod}). \looseness=-1

A second major constraint arises from power allocation. When operating in rank 2 mode, the UE's total transmission power is split between the two data streams. This reduces the power per stream, making it difficult to achieve the high SINR required for higher-order modulation schemes. This limitation is evident in the modulation usage data (Figure \ref{uplink_modulations}). In the urban test, 256QAM usage for UL-MIMO was only 5.2\%, compared to a range of 18.6-21.6\% for UL-CA configurations (Figure \ref{urban_mod}). The disparity widened in the suburban route, where 256QAM usage during UL-MIMO dropped to 1.4\%, while the CA\_n41AA-n25AA configuration maintained a usage of 20.4\% (Figure \ref{suburban_mod}).  In the rural scenario, however, the RF environment became the overriding constraint, forcing all configurations to rely on the lowest-order modulation, with QPSK being used for over 68\% of Resource Blocks (Figure \ref{rural_mod}). \looseness=-1

\section{Conclusions and Future Work} \label{sec_conclusion}

At the time of this research, most commercial networks and UEs do not support UL Tx Switch, necessitating a performance comparison of these mutually exclusive features. Our evaluation on a live 5G NR network demonstrates that UL-CA between a mid-band TDD channel and an FDD channel provides consistently higher and more reliable uplink throughput than UL-MIMO on a single TDD channel. The performance disparity was most significant in low-density environments because poor RF conditions forced the UE to use rank-1 transmit mode, negating the benefits of spatial multiplexing. Conversely, the performance benefits of spatial multiplexing were realized in dense network environments, where the UE transmitted in rank 2 mode for the majority of the time. However, a key limitation for UL-MIMO across all environments was a reduced ability to utilize higher-order modulation schemes, a consequence of splitting the UE's power budget between two spatial streams. \looseness=-1 %The performance disparity was most significant in low-density environments; on the rural route, the median uplink throughput of the CA\_n25AA-n41AA configuration was 115\% greater than that of UL-MIMO. This is because poor RF conditions forced the UE to use the less efficient transmit diversity mode for 89.5\% of transmissions, negating the benefits of spatial multiplexing. Conversely, in the dense urban environment, UL-MIMO operated in the more efficient spatial multiplexing mode 82.5\% of the time, achieving an adequate 10th percentile throughput of 43.4 Mbps. However, a key limitation for UL-MIMO across all environments was a reduced ability to utilize higher-order modulation schemes, a consequence of splitting the UE's power budget between two spatial streams.

%Given these conclusions, operators and manufacturers should prioritize the adoption of UL-CA to allow for overall higher uplink throughput and reliability. This may expand the adoption of current 5G applications, such as FWA and 4K/8K UHD live streaming, as well as encouraging emerging technologies, such as autonomous vehicles, VR/AR, and IoT applications. Operators should seek to densify their networks in order to achieve higher efficiency when UL-MIMO is used, allowing a better experience for UE that may not support UL-CA. Operators may also consider disabling or steering towards single channel UL-MIMO in dense or high traffic environments to avoid assigning resources from multiple channels to one UE. 

Based on these findings, network operators and device manufacturers should prioritize the widespread adoption of UL-CA . To maximize the throughput gains of UL-MIMO, operators should continue to densify networks to ensure favorable RF conditions. Furthermore, operators could implement dynamic traffic steering policies, such as prioritizing single-channel UL-MIMO in congested cells, to optimize spectral efficiency and maximize overall sector capacity by avoiding the allocation of multiple carriers to a single user via UL-CA. \looseness=-1

% In future works, the performance of UL Tx Switch will be evaluated when it is commercially available on networks and UE. The impact of UL-MIMO and UL-CA on live streaming applications, as well as battery life or power consumption will be measured in future research. UL-MIMO in FDD NR bands will also be a focus of study when UE gains support for it.

%Future work will focus on evaluating the performance of Uplink Transmit Switching (UL Tx Switch) once it becomes commercially available, as this technology is designed to utilize both uplink enhancement features simultaneously. Subsequent research will also investigate the practical impact of UL-MIMO and UL-CA on specific applications, such as live streaming, and will quantify their effects on UE power consumption and battery life. Finally, the performance of UL-MIMO on FDD NR bands will become a key area of study as UE support for this feature emerges.

Future work will focus on evaluating the performance of Uplink Transmit Switching (UL Tx Switch) as well as UL-MIMO on FDD NR bands. Subsequent research will also investigate the practical impact of UL-MIMO and UL-CA on specific applications, such as live streaming, and will quantify their effects on UE power consumption and battery life. 

\section*{Acknowledgment}

The authors would like to express their gratitude to \textbf{Pei Xiaohong} of \textit{Qtrun Technologies} for providing \textit{Network Signal Guru (NSG)} and \textit{AirScreen}, the cellular network drive test software used for data collection and analysis in this research. 

\setstretch{0.88}
\renewcommand{\IEEEbibitemsep}{0pt plus 0.5pt}
\makeatletter
\IEEEtriggercmd{\looseness=-1}
\makeatother
\IEEEtriggeratref{1}
\Urlmuskip=0mu plus 1mu\relax

\bibliographystyle{IEEEtran}
\vspace{-0.5mm}
\bibliography{bstcontrol,b_reference}

\end{document}